# Physical representations of corner symmetries

Ludovic Varrin[*]

National Centre for Nuclear Research, Pasteura 7, 02-093 Warsaw, Poland

We give the full representation theory of the gravitational extended corner symmetry group in two dimensions. This includes projective representations, which correspond to representations of the quantum corner symmetry group. We find that they are described by one-dimensional conformal fields with an additional index in the Fock space of the harmonic oscillator. We begin with a review of Mackey's theory of induced representations and then proceed to its application to the corner symmetries. The field representations, induced from the irreducible representations of the special linear group, are worked out first. The little group method is then applied to the extended corner symmetry group to obtain the irreducible unitary representations. Finally, we focus on projective representations and their application to the description of local subsystems.

## I. INTRODUCTION

What are the symmetries of gravity? The response to this seemingly simple question reveals several interesting aspects of spacetime. Gravity can be seen as a gauge theory of the group of diffeomorphisms. While gauge symmetries are usually not considered "physical" symmetries of the theory, some of them turn out to have nonvanishing Noether charges in the presence of boundaries. In particular, those charges have support on codimension-2 surfaces, also called *corners*. While the role of these surface charges started to be understood in the 1970s [1], the modern understanding relies on the concept of *edge modes*. They are physical fields living on the boundary that host a representation of the algebra the surface charges obey. Furthermore, it was recently understood [2] that, in any theory, this algebra turns out to be a subalgebra of some universal one. In particular, for corners located at finite distances, the gravitational charges of any theory will form an subalgebra of

$$\mathfrak{ecs} = \left(\mathfrak{diff}(S)\mathbin{\mathpalette\@loopplus\relax}\mathfrak{sl}(2,\mathbb{R})^S\right)\mathbin{\mathpalette\@loopplus\relax}\left(\mathbb{R}^2\right)^S, \qquad (1)$$

where $S$ denotes the corner and $\mathfrak{ecs}$ refers to the extended corner symmetries algebra. The surprisingly universal character of the algebra, added to the prior recent work on corner symmetries, lead to the *corner proposal* [3–13].

[*]Contact author: ludovic.varrin@ncbj.gov.pl

The proposal is a bottom-up approach to quantum gravity, which uses symmetries as its foundational principle.

This leads us to the next question: why do we care about symmetries? Symmetries play a fundamental role in the entire field of physics, with a particular importance in fundamental physics. In a classical theory, the symmetry group can furnish a complete symplectic structure through the method of coadjoint orbits [14–18]. In a quantum theory, the kinematics are defined by unitary representations of the symmetry group. That is, the symmetries organize the possible physical states of a theory; the most famous example of which is Wigner's classification of the irreducible unitary representations of the Poincaré group [19], which plays a fundamental role in quantum field theory. The corner proposal states that the corner symmetry group should be considered the fundamental component of gravity, just as the Poincaré group is for relativistic field theory. In particular, this implies that the Hilbert space of quantum gravity should host an irreducible unitary representation of some subgroup of the universal corner symmetry group.

While some interesting applications of this idea already exist [20,21], they usually concentrate on the corner diffeomorphisms and special linear part of the extended corner symmetries (ECS), while neglecting the normal translations. However, when using the extended phase space formalism [11], the so-called *embedding edge modes* are used to make the translation charges integrable. In that case, they are therefore on the same footing as the rest of the algebra, and should be considered in a general application of the proposal. Additionally, we acknowledge the existence of similar approaches to the asymptotic version of the symmetry group [22–24], which complement the present work focused on the finite-distance version.

The main barrier to applying directly the proposal resides in the complexity of the algebra in question. The corner diffeomorphisms make the ECS infinite dimensional, and the full representation theory of such groups is, in general, unsolved. One way around this difficulty is to consider, as a first step, only the finite part of the algebra. This was done in our previous work [25], where a concrete application of the corner proposal, which takes normal translations into account, was performed. Note that this can be seen both as a two-dimensional gravity toy model, as well as a per-point analysis of the higher dimensional case. The latter is related to the fact that the full algebra can be seen as a copy of the finite one, for each point of the corner. In some sense, this follows the classical approach of [12] where the orbit method was first applied to the finite part of the ECS and the corner diffeomorphisms were reintroduced later.

One of the main points of [25] was that, when dealing with quantum state representations of the corner symmetries, one needs to consider the existing central extension between the normal translations. This is due to the projective nature of representations in a quantum theory. This leads to the following statement, which will serve as the guiding principle for this paper:

*Physical representations of the extended corner symmetry group in two dimensions are given by irreducible unitary representations of the quantum corner symmetry (QCS) group.*

$$\text{QCS} = \text{SL}(2, \mathbb{R}) \ltimes H_3,$$

where $H_3$ denotes the three-dimensional Heisenberg group. This central extension is of great interest for two main reasons. First, it only appears in the quantum case, as it can be shown that the classical Poisson bracket realization of the gravitational corner symmetries is always without central extensions [11]. Second, this provides a new quantum number labeling the representations, which has natural dimensions of energy squared. While having a dimensionful parameter naturally appearing from a theory is interesting in its own right, it is also believed that it could play an important role in considerations of entanglement entropy.

The second key aspect of our prior work involved the description of local subsystems. The failure of the factorization of Hilbert spaces for local subsystems is a well known fact in gauge theories and gravity. This is due to the gauge constraints, which make the data from one subregion dependent on the data from another. One theoretical solution to this issue was given by Donnelly and Freidel in [3], with the introduction of the entangling product. Our explicit representation of the QCS allowed for the concrete realization of this entangling product. The kinematics of the corner proposal therefore seem well adapted to the issue of localized subsystems in quantum gravity.

The next logical step is to generalize such a construction to the complete representation theory of the corner symmetries. As noted in [25], the full representation theory of the QCS was not known to the authors, and no convincing argument could be made regarding the uniqueness of the "metaplectic" representation employed. The present paper is therefore devoted to establishing the complete mathematical foundations of the representation theory for the finite part of the ECS. As previously mentioned, this comprises the irreducible unitary representations of the QCS. However, since these representations do not reduce to ordinary representations of the ECS in the limit where the central extension goes to zero, we also consider the irreducible unitary representations of the non-centrally extended version. We emphasize that this paper is primarily focused on the mathematical aspect of the problem, whereas most of the physical implications are left for future work.

The symmetry groups of interest are semidirect product groups where the representation theory of both sides of the product is completely known. This is the perfect setting for Mackey's theory of induced representations. While the theory is well known in the physics community, some subtleties, particularly regarding non-Abelian normal subgroups, do not seem to be widely disseminated. Furthermore, it is challenging to find the right balance between mathematical rigor and accessibility for physicists within the existing literature. In the interest of self-contained clarity, we therefore provide a review of the theory, focusing on the material required for the following sections.

The paper is organized as follows. We start with the review of Mackey's theory of induced representations in Sec. II. In addition to serving as a reminder, this section is also used to fix notation. We then move on to the ECS in Sec. III. As it lies at the heart of the corner proposal, we start by giving a brief review of how this group emerges in two-dimensional gravity. Next, we proceed by inducing irreducible representations of the special linear group to the full inhomogeneous group. This serves as a straightforward first application of Mackey's theory presented in the previous section. Furthermore, those representations are in direct analogy with the relativistic fields. We thus call them "field representations." To get irreducible representations, we then move to the "little group" method and obtain the first physical representations of the paper, associated with a null value of the central extension. Finally, Sec. IV is devoted to the representation theory of the QCS. We start by justifying the presence and the form of the extensions and then turn to describing the irreducible unitary representations. We close this section with the generalization of the gluing procedure introduced in [25]. Section V offers comments on the next steps and potential directions for future work.

## II. INDUCED REPRESENTATIONS

The theory of induced representations was developed by Mackey [26–28]. It is a generalization of the famous work of Wigner and Bargmann on the representation theory of the Poincaré group [19,29,30]. Without delving into the mathematical details, we give here a self-contained presentation of the construction of induced representations, with a focus on the case of semidirect groups. We start by giving general considerations on Mackey's theory and then move on to the cases of interest in the present work.

Given a a group $G$, a subgroup $H \subset G$, and a representation of the subgroup on a vector space $V$,

$$D : H \longrightarrow \text{End}(V), \quad h \mapsto D(h), \quad (2)$$

one can construct a representation of the entire group, acting on vector-valued functions on the coset space $\psi : G/H \to V$.[1] In the case of the semidirect group $G = H \ltimes N$, any group element can be decomposed as $g = h \cdot t$ for $h \in H$ and $t \in N$ and the induced representation is given by

$$(U_g \psi)(p) = D(h) \psi(g^{-1} \triangleright p), \quad (3)$$

where "$\triangleright$" denotes the natural action of the group on its coset space.

If $V$ is a Hilbert space and $D$ is unitary with respect to the inner product, one can always make $U$ into a unitary representation. However, even if the representation $D$ is irreducible, the representation $U$ is generally not. In order to obtain irreducible representations, one has to apply the *little group* method, that we will now describe.

The structure of semidirect product groups implies the existence of an automorphism on the normal subgroup given by

$$t \mapsto \tilde{t} = h \cdot t \cdot h^{-1}. \quad (4)$$

Given an irreducible unitary representation of the normal subgroup $\Xi : N \longrightarrow \text{End}(\mathcal{H}^{(\Xi)})$, there exists two subgroups $\mathring{H} \subset H$ and $H^{(0)} \subset H$ and their associated representations,

$$\begin{aligned} \mathring{D} &: \mathring{H} \longrightarrow \text{End}(\mathcal{H}^{(\mathring{D})}), \\ D^{(0)} &: H^{(0)} \ltimes N \longrightarrow \text{End}(\mathcal{H}^{(\Xi)}). \end{aligned} \quad (5)$$

The representation $D^{(0)}$ is an extension of the representation $\Xi$,

$$D^{(0)}|_N = \Xi, \quad (6)$$

which represents the automorphism (4) on the Hilbert space $\mathcal{H}^{(\Xi)}$:

$$\Xi(\tilde{t}) = \Xi(h \cdot t \cdot h^{-1}) = D^{(0)}(h) \Xi(t) D^{(0)}(h)^{-1}. \quad (7)$$

The unitary irreducible representations of the group $G$ can then be induced from representations of the form

$$\mathring{U} = \mathring{D} \otimes D^{(0)}. \quad (8)$$

The two cases of interest in this paper correspond to $H^{(0)}$ being trivial, which, as we will shortly see, arises when $N$ is Abelian, and the case where $H^{(0)} = \mathring{H} = H$.

To treat the Abelian case, we first note that the derivative of the representation $\Xi$ is a representation of the Lie algebra $\mathfrak{n}$. Therefore, given a basis $\{T_a\}$ of $n$, the vectors $\Xi'(T_a)$ span the Hilbert space $\mathcal{H}^{(\Xi)}$. Since the representation $D^{(0)}$ also acts on the Hilbert space of the $\Xi$ representation, the representation $D'^{(0)}$ must be defined in terms of the enveloping algebra of $\mathfrak{n}$. We denote the now-Abelian normal subgroup $N = \mathbb{R}^n$. Since the enveloping algebra is also Abelian, the Lie algebra of $H$ cannot be represented in it and thus the extension $D^0$ is trivial $D^0(h) = 1$ and $H^0$ is the identity element. The representation of the automorphism (7) then reduces to

$$\Xi(\tilde{t}) = \Xi(h \cdot t \cdot h^{-1}) = \Xi(t) \quad \forall \ t \in N. \quad (9)$$

This equation then defines $\mathring{H}$ as the *little group*; that is, the subgroup of $H$ whose automorphism on the normal subgroup $N$ preserves the representation $\Xi$ in the sense of (9). In practice, however, it will be easier to work with the equivalent stabilizer group that we will now define. Since the normal subgroup is a vector space, the semidirect group is defined by a matrix representation of the subgroup $H$,

$$\varphi : H \longrightarrow M_{n \times n}(\mathbb{R}), \quad h \mapsto \varphi[h], \quad (10)$$

and the automorphism (4) can be written

$$\tilde{t} = \varphi[h] t, \quad (11)$$

where $t$ is interpreted as an n-tuple in $\mathbb{R}^n$ and $\varphi[h]$ acts on it via the usual matrix multiplication. Furthermore, the irreducible unitary representations of the Abelian group are one dimensional and $\Xi$ simply becomes the character of the representation. Introducing the dual $\mathbb{R}^{*n}$, and the pairing $\langle \cdot, \cdot \rangle : \mathbb{R}^n \times \mathbb{R}^{*n} \longrightarrow \mathbb{R}$, the elements $p \in \mathbb{R}^{*n}$ label the characters by

$$\Xi_p(t) = e^{i \langle t, p \rangle}. \quad (12)$$

Introducing the dual action $\varphi^*$ on the dual space $\mathbb{R}^{*n}$,

$$\langle \varphi[h] t, p \rangle = \langle t, \varphi^*[h^{-1}] p \rangle, \quad \forall \ t \in \mathbb{R}^n, \quad p \in \mathbb{R}^{*n}, \quad (13)$$

we see from Eqs. (9) and (12) that the little group is isomorphic to the stabilizer,

---

[1]We assumed here the existence of a global section of the bundle $G \xrightarrow{\pi} G/H$ with canonical projection map $\pi$. While this is not a restriction of the general construction, it is true for semidirect groups and will thus suffice in our case.

$$H_p = \{h \in H | \varphi^*[h]p = p\}. \tag{14}$$

We also define the orbit at a point $p \in \mathbb{R}^{*n}$ as

$$\mathcal{O}_p = \{\varphi^*[h]p | h \in H\}. \tag{15}$$

The particularity of the Abelian normal subgroup case hinges on two isomorphisms. First, the orbit $\mathcal{O}_p$ is isomorphic to the coset space $H/H_p$. To identify them, we introduce a map[2]

$$\mathsf{h}: \mathcal{O}_p \longrightarrow H, \qquad q \mapsto \mathsf{h}_q, \tag{16}$$

such that $\varphi^*[\mathsf{h}_q]p = q$ for all $q \in \mathcal{O}_p$. One can then identify a point $q \in \mathcal{O}_p$ with the coset $\mathsf{h}_q H_p \in H/H_p$. The second isomorphism is between the coset space $H/H_p$ and the coset space $G/G_p$, where $G_p = H_p \ltimes N$, which allows us to write the induced representation space as functions on the orbit $\mathcal{O}_p$. We are now finally ready to construct the induced representation. Given a unitary representation of the little group $\mathring{D}: H_p \longrightarrow \text{End}(\mathring{\mathcal{H}})$, we construct the representation $D: G_p \longrightarrow \text{End}(\mathring{\mathcal{H}})$ defined as

$$D(\mathring{h}, t) = \Xi_p(t)\mathring{D}(\mathring{h}) = e^{i\langle t, p \rangle}\mathring{D}(\mathring{h}), \tag{17}$$

for $\mathring{h} \in H_p$ and $t \in \mathbb{R}^n$. Using the isomorphisms described above, we can finally write the induced representation (3) of $D$ to the entire group $G$ as

$$(U_{(h,t)}\psi)(q) = e^{i\langle t, q \rangle}\mathring{D}(\mathsf{h}_q^{-1} \cdot h \cdot \mathsf{h}_{\varphi^*[h^{-1}]})\psi(\varphi^*[h^{-1}]q), \tag{18}$$

where $q \in \mathcal{O}_p$. As a result, each distinct orbit will have its own set of representations. In the Poincaré case, this is what leads to the different spin and mass states. Note that it can be shown that the representation (18) does not depend on the choice of orbit representative. If the representation $\mathring{D}$ is irreducible, Mackey theory ensures that the induced representations (18) form a complete set of irreducible unitary representations of $G$. We also note that the intimate relationship between orbits and irreducible representations of a group go beyond the semiproduct case. This is the subject of the coadjoint orbit method of Kirillov [14–18].

If the normal subgroup is not Abelian, its representations are in general more complicated than the simple character (12). However, the case of interest in this paper is captured by the following specific instance. If the representation $D^0$ is defined globally, that is $G^0 = G$, then by Eq. (7), the little group is the entire group $H$. The representations (8) are then already representations of the entire group and Mackey's theory ensures they cover all possible unitary irreducible

---

[2]It can be shown that this map is well defined and unique up to right composition by any function that sends $\mathcal{O}_p$ to $H_p$.

---

representations. They are given explicitly by

$$U_{(h,t)} = D(h) \otimes D^{(0)}(h)\Xi(t), \tag{19}$$

where $D: H \longrightarrow \text{End}(\mathcal{H}^{(D)})$ and $D^{(0)}: H \longrightarrow \text{End}(\mathcal{H}^{(\Xi)})$ are unitary irreducible representation of $H$. For more details, see for example [31].

## III. EXTENDED CORNER SYMMETRY GROUP

The symmetry group of gravity consists of the diffeomorphisms on the spacetime manifold. Although these are usually understood as gauge symmetries, the presence of finite distance boundaries makes a subgroup of them physical and located on the boundary. The algebra of that subgroup is called the *universal corner algebra*. In order to understand how it arises, we start by giving the two-dimensional analog of the calculation performed in [2].

### A. Corner symmetries of two-dimensional gravity

We consider a two-dimensional spacetime $M$ and focus on a corner, simply defined by a point $p_S$, embedded in the spacetime by the trivial embedding,

$$\phi_0: p_S \to M. \tag{20}$$

This particular embedding corresponds to the choice of local coordinates $y_0^M(p) = \{x^a, a = 1, 2\}$, in which the corner point is located at the origin. Given two vector fields $\xi, \chi \in TM$, one can thus expand them near the corner in the trivial embedding coordinates,

$$\xi = \xi^a(x)\partial_a = (\xi_{(0)}^a + x^b \xi_{(1)b}^a)\partial_a + \mathcal{O}(x^2), \tag{21}$$

$$\chi = \chi^a(x)\partial_a = (\chi_{(0)}^a + x^b \chi_{(1)b}^a)\partial_a + \mathcal{O}(x^2). \tag{22}$$

Their commutator at $\mathcal{O}(x)$ is then given by

$$\begin{aligned}[\xi, \chi] &= [(\xi_{(0)}^a + x^b \xi_{(1)b}^a)\chi_{(1)a}^c - (\chi_{(0)}^a + x^b \chi_{(1)b}^a)\xi_{(1)a}^c]\partial_c \\ &= [\chi_{(1)a}^c \xi_{(0)}^a - \xi_{(1)a}^c \chi_{(0)}^a]\partial_c + x^b[\chi_{(1)a}^c \xi_{(1)b}^a - \xi_{(1)a}^c \chi_{(1)b}^a]\partial_c \\ &= [\chi_{(1)a}^c \xi_{(0)}^a - \xi_{(1)a}^c \chi_{(0)}^a]\partial_c - x^b[\xi_{(1)}, \chi_{(1)}]_b^c \partial_c.\end{aligned} \tag{23}$$

The first line corresponds to the general linear group acting on normal translations and the second line is simply the general linear algebra. Thus, in the case of two dimensions, the *universal corner symmetry group* is given by

$$\text{UCS} = \text{GL}(2, \mathbb{R}) \ltimes \mathbb{R}^2. \tag{24}$$

In order to see how only the traceless part of the general linear group matters for corners located at a finite distance, let us look at the example of Einstein-Hilbert theory. We first introduce a metric

$$ds^2 = h_{ab}(x)dx^a dx^b. \tag{25}$$

Since the corner is not at infinity, we can also expand the above near the corner point,

$$h_{ab}(x) = h_{ab}^{(0)} + x^c h_{abc}^{(1)} + \mathcal{O}(x^2). \tag{26}$$

In the Einstein-Hilbert theory, the diffeomorphism charge is given by [3]

$$H_\xi = Q_\xi|_{p_S} = \phi_0^*(\star dh(\xi, \cdot)), \tag{27}$$

where the star denotes the Hodge dual, and with

$$\begin{aligned} dh(\xi, \cdot) &= \frac{1}{2}\partial_c(h_{ab}\xi^a)dx^b \wedge dx^c \\ &= \frac{1}{2}(h_{ab,c}\xi^a + h_{ab}\xi^a_{,c})dx^b \wedge dx^c \\ &= \frac{1}{2}(h_{abc}^{(1)}\xi^a_{(0)} + h_{ab}^{(0)}\xi^a_{(1)c} + x^d(\cdots)_{bcd} \\ &\quad + \mathcal{O}(x^2))dx^b \wedge dx^c, \end{aligned} \tag{28}$$

where the dots denote a constant term. The Hodge dual of the above term, pulled back to the corner, gives

$$\phi_0^*(\star dh(\xi, \cdot)) = \frac{1}{2}\sqrt{-h^{(0)}}\left(h_{abc}^{(1)}\xi^a_{(0)} + h_{ab}^{(0)}\xi^a_{(1)c}\right)\epsilon^{cb}. \tag{29}$$

The charge associated to diffeomorphisms is therefore

$$H_\xi = (\xi^a_{(0)}p_a + \xi^a_{(1)c}N^c_a), \tag{30}$$

where we defined

$$N^c_a = \frac{1}{2}h_{ab}^{(0)}\epsilon^{bc}, \tag{31}$$

$$p_a = \frac{1}{2}h_{abc}^{(1)}\epsilon^{bc}. \tag{32}$$

In the above equations, the constant factor $\sqrt{-h^{(0)}}^{-1}$ was dropped since it does not change the algebra. Note that the term defined in (31) is traceless, which means that only the traceless part of $\xi^a_{(1)c}$ will contribute in (30). Thus, only the $SL(2,\mathbb{R})$ subgroup of the general linear group is activated in the charge algebra. In the case of finite distance corners, the corner symmetry group therefore reduces to

$$ECS = SL(2,\mathbb{R}) \ltimes \mathbb{R}^2, \tag{33}$$

where $ECS$ stands for *extended corner symmetry*. The associated lie algebra, $\mathfrak{ecs}$, is spanned by five generators $L_0, L_\pm, P_\pm$ where the first three generate the $\mathfrak{sl}(2,\mathbb{R})$ algebra

$$[L_-, L_+] = 2L_0, \quad [L_0, L_\pm] = \pm L_\pm, \tag{34}$$

and the Abelian translations $P_\pm$ have the following cross-commutation relations:

$$[L_0, P_\pm] = \pm\frac{1}{2}P_\pm, \quad [L_\pm, P_\mp] = \mp P_\pm, \quad [L_\pm, P_\pm] = 0. \tag{35}$$

The algebra admits one cubic Casimir [12]

$$\mathcal{C}_{ECS} = L_+ P_-^2 + L_- P_+^2 - 2L_0 P_+ P_-. \tag{36}$$

In direct analogy with the Poincaré group in quantum field theory, the corner proposal can be stated as the following: In the two-dimensional case, the states of quantum gravity must fall into a physical[3] representation of the ECS. One can thus see the importance of classifying all representations of the corner symmetry group, as it corresponds to the kinematics of two-dimensional quantum gravity, from the corner proposal point of view.

### B. Field representations

Our application of the formalism presented in the previous section to the corner symmetry groups starts by inducing irreducible representations of $SL(2,\mathbb{R})$ to the entire ECS group. Although such representations are not generally unitary, they remain highly relevant in physics. Indeed, this is the case of relativistic fields which generally are in nonunitary representations of the Poincaré group. In fact, they can be obtained by inducing the finite dimensional irreducible representations of the Lorentz group, described using the local isomorphism to $SU(2) \times SU(2)$, to the full inhomogeneous group by the method detailed in the previous section. Since the Lorentz group is noncompact, those finite dimensional representations cannot be unitary. In this section, we will apply the same analysis to the ECS group. The special linear group being locally isomorphic to $SU(2)$, we can construct the associated finite dimensional irreducible representations and induce them to the entire corner symmetry group using Mackey's theory. This will serve as a straightforward application of the theory of induced representations and a good warm-up for the more complicated cases ahead. As previously mentioned, the induced representations will not generally be irreducible. In order to obtain irreducible representations, one needs to apply the little group method, which will be the topic of the next section.

Since $\mathfrak{sl}(2,\mathbb{R}) \cong \mathfrak{su}(2)$, the finite dimensional representation theory of $SL(2,\mathbb{R})$ is equivalent to the representations of $SU(2)$. To make this isomorphism more explicit, let us introduce the angular momentum basis of the algebra,

---

[3]What is meant by physical representation will be clarified in Sec. IV.

$$J_0 = 2L_0, \quad J_+ = L_+, \quad J_- = -L_-. \tag{37}$$

For any $j \in \mathbb{N}$ or $j \in \frac{\mathbb{N}}{2}$ there exists a $2j+1$ dimensional representation spanned by the eigenvectors of $J_0$,

$$J_0^{(j)} \psi_m^{(j)} = m \psi_m^{(j)}, \tag{38}$$

where $m = -j, -j+1, \ldots, j$ and the $(j)$ subscript indicates that the generator is taken in the corresponding representation. The remaining generators of $\mathfrak{sl}(2, \mathbb{R})$ then act as raising and lowering operators,

$$J_\pm^{(j)} \psi_m^{(j)} = \sqrt{j(j+1) - (m(m \pm 1))} \psi_{m \pm 1}^{(j)}, \tag{39}$$

such that $J_-^{(j)} \psi_{-j}^{(j)} = J_+^{(j)} \psi_j^{(j)} = 0$. The Hilbert spaces carry a representation of the group by exponentiation of the generators: Any element $h \in \mathrm{SL}(2, \mathbb{R})$ can be written as

$$h = \exp(h^i J_i), \tag{40}$$

where $h^i, i = 0, -, +$ are coordinates on the Lie group.[4] The group representation can then be written

$$D^{(j)}(h) \psi_m = \sum_{k=0}^{\infty} \frac{(h^i J_i^{(j)})^k}{k!} \psi_m, \quad \forall\, h \in \mathrm{SL}(2, \mathbb{R}), \tag{41}$$

where the $j$ index of the vectors was dropped for simplicity.

We now construct the induced representations of the ECS. In order to do so, we first observe that the coset space is isomorphic to $\mathbb{R}^2$:

$$\mathrm{ECS}/\mathrm{SL}(2, \mathbb{R}) \cong \mathbb{R}^2. \tag{42}$$

The induced representation will therefore be constructed on vector-valued functions,

$$\psi \colon \mathbb{R}^2 \longrightarrow \mathcal{H}^{(j)}, \tag{43}$$

where $\mathcal{H}^{(j)}$ denotes the Hilbert space of the $j$ representation. Any element $g \in \mathrm{ECS}$ can be written as $g = t \cdot h = \exp(t^a P_a) \exp(h^i J_i)$, where $a = -, +$. The induced representation (3) can then be written

---
[4]Since $\mathrm{SL}(2, \mathbb{R})$ is not simply connected, the exponentiation of an algebra element will be an element of the universal cover of the special linear group $\widetilde{\mathrm{SL}(2, \mathbb{R})}$. However, the fact that the maximal compact subgroup has integer values ($e^{2\pi i L_0} = \pm 1$) is exactly the condition needed for the representations to be of $\mathrm{SL}(2, \mathbb{R})$ rather than its universal cover.

$$(U_{(t \cdot h)}^{(j)} \psi_m)(x) = D^{(j)}(h) \psi_m \big( \varphi[\exp(-h^i J_i)](x - t) \big)$$
$$= \sum_{k=0}^{\infty} \frac{(h^i J_i^{(j)})^k}{k!} \psi_m \big( \varphi[\exp(-h^i J_i)](x - t) \big), \tag{44}$$

where we have used that the action of the group on the coset space is simply given by

$$(t \cdot h) \triangleright x = \varphi[h] x + t. \tag{45}$$

For any irreducible $j$ representation of $\mathrm{SL}(2, \mathbb{R})$, Eq. (44) defines a representation of the ECS group. Let us now compute the representation of the algebra:

$$(U'_X \psi_m)(x) = \frac{\mathrm{d}}{\mathrm{d}\alpha} \Big( (U_{\exp(\alpha X)}^{(j)} \psi_m)(x) \Big) \big|_{\alpha = 0}, \tag{46}$$

where $X \in \mathfrak{ecs}$. The translations act as derivative operators,

$$(U'_{P_\pm} \psi_m)(x) = -\frac{\partial \psi_m(x)}{\partial x_\pm}, \tag{47}$$

and the $\mathfrak{sl}(2, \mathbb{R})$ generators act as

$$(U'_{J_i} \psi_m)(x) = J_i^{(j)} \psi_m(x) + (\varphi'[J_i])^a{}_b x^b \partial_a \psi_m(x), \tag{48}$$

where $\varphi' \colon \mathfrak{sl}(2, \mathbb{R}) \longrightarrow M_{2 \times 2}(\mathbb{R})$ is the matrix representation of the Lie algebra defined as

$$\varphi'[X] = \frac{\mathrm{d}}{\mathrm{d}\alpha} (\varphi[-\alpha X]). \tag{49}$$

To write this action more explicitly, let us introduce the explicit matrix representation,

$$\varphi'[J_0] = \begin{pmatrix} 1 & 0 \\ 0 & -1 \end{pmatrix}, \quad \varphi'[J_+] = \begin{pmatrix} 0 & 1 \\ 0 & 0 \end{pmatrix},$$
$$\varphi'[J_-] = \begin{pmatrix} 0 & 0 \\ 1 & 0 \end{pmatrix}. \tag{50}$$

We get

$$(U'_{J_0} \psi_m)(x) = m \psi_m(x) + (x_- \partial_{x_-} - x_+ \partial_{x_+}) \psi_m(x),$$
$$(U'_{J_+} \psi_m)(x) = \sqrt{j(j+1) - m(m+1)} \psi_{m+1}(x)$$
$$\quad + x_- \partial_{x_+} \psi_m(x)$$
$$(U'_{J_-} \psi_m)(x) = \sqrt{j(j+1) - m(m-1)} \psi_{m-1}(x)$$
$$\quad + x_+ \partial_{x_-} \psi_m(x). \tag{51}$$

This action is reminiscent of the action of the Lorentz algebra on the fields of a relativistic field theory. The first terms are the "spin" parts, corresponding to a transformation of the fields in some internal space, and the inhomogeneous terms are the "orbital" parts corresponding to the actual spacetime transformations. In particular, note that if we choose the trivial representation $D^{(j=0)}$, only the inhomogeneous part remains. The thus obtained representation is called the defining representation in [12]. It has a vanishing Casimir value and is irreducible. In the analogy with the Poincaré group, these would correspond to scalar fields. However, as expected, the representation is not irreducible for a general $j$. To see this, one can use the induced representation of the algebra to compute the action of the Casimir operator (36) on an arbitrary $j$ representation and find that it is not a multiple of the identity. Furthermore, since the representations of $SL(2,\mathbb{R})$ from which we induced are not unitary, neither are the resulting representations of the ECS. One could instead induce the infinite dimensional unitary representations of the special linear group discussed in Sec. IV. Although they would have a very similar form to (44), the unbounded nature of the $j$ index would allow for unitarity. These representations are reminiscent of the "expinor" theory of Harish-Chandra and Dirac [32,33] (see also [34] for a historical review), which does not have any known physical relevance.

### C. Irreducible representations

In order to apply the little group method to the ECS, we need to consider the orbits of $SL(2,\mathbb{R})$ in $\mathbb{R}^{*2} \cong \mathbb{R}^2$. To do so, let us introduce the *KAN* matrix representation of the special linear group. Any element $g \in SL(2,\mathbb{R})$ can be written uniquely as

$$\varphi(g) = \begin{pmatrix} \cos(\theta) & -\sin(\theta) \\ \sin(\theta) & \cos(\theta) \end{pmatrix} \begin{pmatrix} \frac{1}{r} & 0 \\ 0 & r \end{pmatrix} \begin{pmatrix} 1 & x \\ 0 & 1 \end{pmatrix}. \quad (52)$$

The parameters $\theta \in [0, 2\pi]$, $r \geq 0$ and $x \in \mathbb{R}$ are thus coordinates on the group and we will denote the group element defined by its *KAN* decomposition by

$$g_{\theta,r,x} = k_\theta \cdot a_r \cdot n_x. \quad (53)$$

Let us now describe the orbits. Any point $(p_-, p_+) \in \mathbb{R}^2$, which is different from the origin, can be reached from the point (1,0) by first scaling the element to the right size, and then rotating it to the right vector,

$$\begin{pmatrix} p_- \\ p_+ \end{pmatrix} = \varphi(k_{\tilde{\theta}}) \varphi(a_{\tilde{r}}) \begin{pmatrix} 1 \\ 0 \end{pmatrix}, \quad (54)$$

for any $(p_-, p_+) \neq (0,0)$ with

$$\tilde{r} = \sqrt{p_-^2 + p_+^2}, \quad \tilde{\theta} = \text{sgn}(p_+) \arccos\left(\frac{p_-}{\sqrt{p_-^2 + p_+^2}}\right). \quad (55)$$

The above construction shows that there exists only two distinct orbits of the action of $SL(2,\mathbb{R})$ on $\mathbb{R}^2$: the nontrivial orbit $\mathcal{O}_{(1,0)}$ and the trivial one containing only the origin $\mathcal{O}_0$. Since the latter corresponds to a trivial action of the translations, we will only look at the former. Sticking with (1,0) as the orbit representative, it is easy to see that the little group is simply the $N$ component of the *KAN* decomposition,

$$H_{(1,0)} = \{n_x | x \in \mathbb{R}\} \cong \mathbb{R}. \quad (56)$$

Since the group is Abelian, its irreducible unitary representations are once again fully described by its characters

$$\mathring{D}^{(\nu)}: H_{(1,0)} \longrightarrow \mathbb{C}, \quad \mathring{D}^{(\nu)}(n_x) = e^{i\nu x}, \quad (57)$$

where the distinct representations are labeled by the different values of $\nu \in \mathbb{R}$. We can now construct the induced representation which acts on functions on the orbit. To emphasize that we are now working in the state representation, we will denote these functions by $\psi(p) \equiv |p\rangle$, where $p = (p_-, p_+) \neq (0,0)$. For a ECS group element $g = t \cdot h$, the induced representation (18) is given by

$$U^{(\nu)}_{(t,h)}|p\rangle = e^{i\langle t,p\rangle} \mathring{D}^{(\nu)}(\mathsf{h}_p^{-1} \cdot h \cdot \mathsf{h}_{\varphi^*[h^{-1}]p})|\varphi^*[h^{-1}]p\rangle, \quad (58)$$

where we recall that $\mathsf{h}_p$ is the $SL(2,\mathbb{R})$ group element that brings the representative to the point $p$. In this case, it is therefore simply given by

$$\mathsf{h}_p = k_{\tilde{\theta}} \cdot a_{\tilde{r}}, \quad (59)$$

with $\tilde{\theta}$ and $\tilde{r}$ given by Eq. (55). Since we chose irreducible unitary representations of the little group, Mackey's theory ensures that one can always equip the Hilbert space with a scalar product such that the induced representation is unitary. Equation (58) then describes the complete set of irreducible unitary representations of the ECS group. The representations are characterized by a single parameter $\nu \in \mathbb{R}$, reflecting the existence of a unique Casimir operator. In fact, one can show that $\nu$ is precisely the scalar value that $\mathcal{C}_{\text{ECS}}$ takes on these representations. In order to do so, we can use the explicit matrix representation of the $\mathfrak{sl}(2,\mathbb{R})$ generators (50).[5] The induced representation of the algebra can then be calculated using elementary matrix calculations. For the $\mathfrak{sl}(2,\mathbb{R})$ generators we get

---

[5]Since we are now working with unitary representations, we choose the matrices of (50) to define the generators multiplied by the imaginary unit. This ensures that the corresponding operators $U'^{(\nu)}(L_i)$ are Hermitian.

$$U'^{(\nu)}(L_0)|p\rangle = \left[2\nu\frac{p_-p_+}{(p_-^2+p_+^2)^2} + \frac{1}{2}(p_+\partial_{p_+} - p_-\partial_{p_-})\right]|p\rangle,$$

$$U'^{(\nu)}(L_-)|p\rangle = \left[\nu\frac{p_-^2-p_+^2}{(p_-^2+p_+^2)^2} + p_-\partial_{p_+}\right]|p\rangle,$$

$$U'^{(\nu)}(L_+)|p\rangle = \left[\nu\frac{p_+^2-p_-^2}{(p_-^2+p_+^2)^2} - p_+\partial_{p_-}\right]|p\rangle. \quad (60)$$

The translations simply act by multiplication,

$$P_\pm|p\rangle = ip_\pm|p\rangle. \quad (61)$$

Choosing the trivial representation of the little group ($\nu = 0$) and applying a Fourier transform, one obtains again the defining representation of the ECS. Finally, it is straightforward to check that

$$U'^{(\nu)}(\mathcal{C}_{\text{ECS}})|p\rangle = \nu|p\rangle. \quad (62)$$

We conclude this section with a remark about the relationship between the field representation of the previous section and the state representation that was just constructed. In the context of quantum field theory, the field representation is related to the state representation by introducing field operators acting on the Hilbert space of the latter, $\hat{\psi}_m(x)$, which transform according to

$$(U_{(t,h)}^{(\nu)})^{-1}\hat{\psi}_m(\varphi[h]x + t)U_{(t,h)}^{(\nu)} = D^{(j)}(h)\hat{\psi}_m(x), \quad (63)$$

where $D^{(j)}$ is the representation defined in Eq. (41).

## IV. QUANTUM CORNER SYMMETRY GROUP

In the previous section we gave the complete description of the irreducible unitary representations of the ECS group. However, they form only a small subset of what we call the physical representations. A quantum theory is defined by a Hilbert space $\mathcal{H}$ and the transition probabilities between different states,

$$\mathcal{P}_{\alpha\to\beta} = |\langle\alpha|\beta\rangle|^2, \quad (64)$$

where $|\alpha\rangle, |\beta\rangle \in \mathcal{H}$ are vectors in the Hilbert space, and the bra-ket denotes the scalar product. An immediate consequence is that the theory's predictions do not depend on the vector phases. This fact is implemented by defining the states of the theory as the *ray operators* rather than the vectors themselves. The former are defined as equivalence classes of vectors, with the following equivalence relation:

$$|\alpha\rangle \sim e^{i\theta}|\alpha\rangle \quad (65)$$

for any $\theta \in \mathbb{R}$. That is, vectors that only differ by a phase are different representative elements of the same ray operator. The states are then elements of the projective Hilbert space

$$P\mathcal{H} = \mathcal{H}/\sim. \quad (66)$$

Such a Hilbert space can be obtained from a symmetry group by considering the projective representations of the latter. A representation $U: G \longrightarrow \text{End}(P\mathcal{H})$ is called projective, if the following equation is satisfied:

$$U_{g_1}U_{g_2} = \omega(g_1, g_2)U_{g_1 \cdot g_2}, \quad (67)$$

where $\omega(g_1, g_2)$ are the two-cocycles of the group. Because of a famous theorem by Bargmann and Mackey [35–37], it is known that the projective representations of a path-connected Lie group are equivalent to the ordinary unitary representations of its maximal central extension. The "maximal" encapsulates two different origins of the central extensions. The first one comes from the first homotopy group and is linked with the universal cover of the symmetry group. In the most known example of Poincaré symmetry, this is the reason why we study representations of $\text{SL}(2,\mathbb{C}) \ltimes \mathbb{R}^4$ instead of the original Poincaré group with Lorentz symmetries. The second one is the addition of a central generator determined by the two-cocycles of the algebra. Unlike Poincaré symmetries, which do not allow nontrivial two-cocycles, the Galilean group—the symmetry group of nonrelativistic mechanics—does. The added central generator is then identified with the mass of the nonrelativistic particles. This last example highlights the strong physical significance of the projective nature of the representations. The unitary irreducible representations of the corner symmetry group that describe a quantum system are therefore those of its maximally centrally extended version, which we discuss in the next section.

### A. Central extensions and Casimirs

The maximal central extension of a group includes contributions from both the first homotopy group and the second cohomology group. We start by addressing the former while the latter will be discussed at the end of the section. The second cohomology group of an algebra is given by the set of two-cocycles modulo trivial ones. Let us denote the generators of the $\mathfrak{ecs}$ by $X_i$, $i = 1, \ldots, 5$. We have

$$[X_i, X_j] = C_{ij}^k X_k, \quad (68)$$

where $C_{ij}^k$ are the structure constants of the algebra. The possible central extensions are then described by two-cocycles $B: \mathfrak{ecs} \otimes \mathfrak{ecs} \longrightarrow \mathfrak{u}(1)$ valued in the Abelian Lie algebra of the unitary group. They enter the extended commutation relations as

$$[X_i, X_j]_B = C_{ij}^k X_k + B(X_i, X_j). \quad (69)$$

In order to describe an algebra extension, the cocycle must obey two properties. First, the cocycle is antisymmetric, which comes from the antisymmetricity of the Lie bracket.

Second, the Jacobi identity of the algebra imposes the following conditions on the cocycle:

$$C^p_{jk}B_{ip} + C^p_{ki}B_{jp} + C^p_{ij}B_{kp} = 0, \qquad (70)$$

where $B_{ij} = B(X_i, X_j)$. A two-cocycle is called trivial if it can be reabsorbed into the definition of the structure constants; that is, if there exists coefficients $\lambda_k \in \mathbb{R}$ such that

$$B_{ij} = C^k_{ij}\lambda_k. \qquad (71)$$

Determining the second cohomology group of the algebra is therefore equivalent to finding all antisymmetric matrices $B_{ij}$ satisfying relation (70), while retaining only the nontrivial ones. Using the explicit structure constants of the $\mathfrak{ecs}$, one finds that the only nontrivial two-cocycle corresponds to a central extension between the translations,

$$[P_-, P_+] = C, \qquad (72)$$

where we have dropped the $B$ subscript of the commutator for simplicity. This extension transforms the translational part of the ECS into the three-dimensional Heisenberg group. The resulting group was first introduced in [25] and is called the *quantum corner symmetry* (QCS) group.[6]

In order to obtain the maximally centrally extended group, one further needs to consider the universal covering of the special linear part. The physical representations of the ECS group are therefore equivalent to the irreducible unitary representations of

$$\widetilde{\text{QCS}} = \widetilde{\text{SL}(2, \mathbb{R})} \ltimes H_3, \qquad (73)$$

where the tilde denotes the universal cover, and $H_3$ is the three-dimensional Heisenberg group.

Let us remark on these central extensions. The extension arising from the cocycles is reminiscent of the Galilean mass operator, appearing as the central element extending the commutator between translations and Galilean boosts. In that case, the extension is necessary to address massive nonrelativistic particles in both the quantum and classical frameworks. In the present case, however, it was shown in [11] that, within the extended phase space formalism, the covariant phase space always realizes the corner symmetries without any nontrivial cocycles. The appearance of the central element (72) is therefore a purely quantum effect. The story is different for the universal cover. The extended corner symmetry group was derived at the algebra level first (cf. Sec. III). Since $\text{SL}(2, \mathbb{R})$ is not simply connected, there is no one-to-one correspondence between the ECS group and its algebra. However, the unique simply connected group that can be associated with the $\mathfrak{ecs}$ is

$$\widetilde{\text{ECS}} = \widetilde{\text{SL}(2, \mathbb{R})} \ltimes \mathbb{R}^2. \qquad (74)$$

This suggests that the universal cover of the special linear group should already be considered at the classical level.

We conclude this section by noting that the QCS possesses two Casimirs, reflecting the fact that the maximal coadjoint orbits are four dimensional. The central element, $C$, is trivially a Casimir. The second Casimir can be obtained by solving a set of partial differential equations on the coadjoint orbits, and can be written as

$$\mathcal{C}_{\text{QCS}} = \frac{C}{2}\left(L_0 + \frac{3}{8}\right) - C\mathcal{C}_{\text{SL}(2,\mathbb{R})} - \frac{1}{2}\mathcal{C}_{\text{ECS}}, \qquad (75)$$

where

$$\mathcal{C}_{\text{SL}(2,\mathbb{R})} = -L_0^2 + \frac{1}{2}(L_+L_- + L_-L_+) \qquad (76)$$

is the quadratic Casimir of the special linear algebra. As was already mentioned in [25], the presence of the central element makes it so we can always shift the value of the cubic Casimir with a simple redefinition. The specific factors in (75) were chosen for later convenience.

### B. Physical representations

The complete classification of the quantum states associated with the corner symmetries has reduced to the full description of the irreducible unitary representations of the group (73). This section is devoted to the construction of these representations. We refer the mathematically inclined readers to [31,38–40]. For notational clarity, operators in a certain representation will, from now on, be denoted by the operator itself, with the indices defining the representation in subscript.

We start with the representations of the normal subgroup $H_3$. In an irreducible representation, the central element $C$ must act as a multiple of the identity. Once the coefficient $c \in \mathbb{R}$ is fixed, Stone–von Neumann's theorem [41] ensures that there exists a unique unitary representation,

$$\Xi^{(c)}: H_3 \longrightarrow \text{End}(\mathcal{F}), \qquad (77)$$

where $\mathcal{F}$ can be taken to be the Fock space of the quantum harmonic oscillator. The representation of the algebra can then be written

$$P_-^{(c)} = \sqrt{c}a, \quad P_+^{(c)} = \sqrt{c}a^\dagger, \quad C^{(c)} = c, \qquad (78)$$

where $a, a^\dagger$ are the usual creation and annihilation operator

$$[a, a^\dagger] = 1. \qquad (79)$$

---

[6]In the mathematical literature, this structure is known as the Jacobi group [38].

In (78) (and in what follows) we assumed $c > 0$. Although the negative case can be treated in a similar fashion, it is important to note that it gives rise to a nonequivalent set of representations. The Weil representation of the special linear algebra,

$$\mathring{D}': \mathfrak{sl}(2, \mathbb{R}) \longrightarrow \text{End}(\mathcal{F}), \qquad (80)$$

expresses the generators of $\mathfrak{sl}(2, \mathbb{R})$ in the universal algebra of the Heisenberg subalgebra

$$\mathring{L}_0 = \frac{1}{2}\left(a^\dagger a + \frac{1}{2}\right), \quad \mathring{L}_- = \frac{1}{2}aa, \quad \mathring{L}_+ = \frac{1}{2}a^\dagger a^\dagger. \qquad (81)$$

This was called the metaplectic representation in [25]. By exponentiation, (81) becomes a global representation of $\widetilde{\text{SL}(2, \mathbb{R})}$ that obeys condition (7).[7] Therefore, Mackey's theory ensures [cf. (19)] that the complete set of irreducible unitary representations of the $\widetilde{\text{QCS}}$ are of the form

$$U^{(\Delta,c)}: \widetilde{\text{QCS}} \longrightarrow \text{End}(\mathcal{H}^{(\Delta)} \otimes \mathcal{F}),$$
$$U^{(\Delta,c)}_{(h,t)} = D^{(\Delta)}(h) \otimes \mathring{D}(h)\Xi^{(c)}(t), \qquad (82)$$

where $D^{(\Delta)}: \widetilde{\text{SL}(2, \mathbb{R})} \longrightarrow \text{End}(\mathcal{H}^{(\Delta)})$ are irreducible unitary representations of the universal cover of the special linear group. Following [42], the later are completely defined through the value of the group element corresponding to the maximal compact subgroup $e^{2\pi i L_0}$ which takes value $e^{-2\pi i \epsilon}$ where $\epsilon \in \mathbb{R}$, and the parameter $\Delta$ that is related to the Casimir operator through $\mathcal{C}^{(\Delta)}_{\text{SL}(2,\mathbb{R})}|\psi\rangle = \Delta(1-\Delta)|\psi\rangle$ for any state $|\psi\rangle$. The representation is then written on eigenstates of $L_0$ as

$$L_0|n\rangle = -n|n\rangle,$$
$$L_-|n\rangle = -\sqrt{(n+\Delta)(n+1-\Delta)}|n+1\rangle,$$
$$L_+|n\rangle = -\sqrt{(n-\Delta)(n-1+\Delta)}|n-1\rangle. \qquad (83)$$

Note that the minus sign convention in the first line implies that $L_+$ ($L_-$) lowers (raises) the eigenvalue by 1. We also note that, at this point, $n$ is not necessarily an integer. We will talk in more detail about the possible values of $n$ and $\Delta$ when we discuss unitarity.

Using Eq. (82), the above representations, together with the Weil representation [Eqs. (78) and (81)], describe the complete set of irreducible representations of the $\widetilde{\text{QCS}}$. They are labeled by two parameters reflecting the existence of two Casimirs. The central element is directly related to the parameter $c \in \mathbb{R}_{>0}$. To understand the role of the cubic Casimir (75), let us write the explicit algebra representation. For a vector $X = L + P \in \mathfrak{qcs}$, it is simply given by

$$U'^{(\Delta,c)}(X) = D'^{(\Delta)}(L) \otimes \mathring{D}'(L)\Xi'^{(c)}(P). \qquad (84)$$

Let us denote the elements of $\mathcal{H}^{(\Delta)} \otimes \mathcal{F}$ by $|n\rangle \otimes |k\rangle \equiv |n, k\rangle$. We then have

$$P^{(\Delta,c)}_-|n, k\rangle = \sqrt{c}\sqrt{k}|n, k-1\rangle, \qquad (85)$$

$$P^{(\Delta,c)}_+|n, k\rangle = \sqrt{c}\sqrt{k+1}|n, k+1\rangle, \qquad (86)$$

$$L^{(\Delta,c)}_0|n, k\rangle = \left(\frac{1}{2}k + \frac{1}{4} - n\right)|n, k\rangle, \qquad (87)$$

$$L^{(\Delta,c)}_-|n, k\rangle = -\sqrt{(n+\Delta)(n+1-\Delta)}|n+1, k\rangle$$
$$+ \frac{1}{2}\sqrt{k(k-1)}|n, k-2\rangle, \qquad (88)$$

$$L^{(\Delta,c)}_+|n, k\rangle = -\sqrt{(n-\Delta)(n-1+\Delta)}|n-1, k\rangle$$
$$+ \frac{1}{2}\sqrt{(k+1)(k+2)}|n, k+2\rangle. \qquad (89)$$

It then simply follows that the action of the cubic Casimir gives

$$\mathcal{C}^{(\Delta,c)}_{\text{QCS}}|n, k\rangle = c\Delta(1-\Delta)|n, k\rangle. \qquad (90)$$

Remarkably, the value of the cubic QCS Casimir in the irreducible representations is given by the value of the corresponding $\text{SL}(2, \mathbb{R})$ Casimir, multiplied by the scalar value of the central element of the Heisenberg algebra. We note that, choosing the trivial representation of $\widetilde{\text{SL}(2, \mathbb{R})}$, we recover the representation used in [25]. In the present conventions, this corresponds to a vanishing value of the cubic Casimir.

Let us now comment on unitarity. The complex structure compatible with the commutation relations (34), (35), and (72) (and the Hermiticity of the Casimir operators) is

$$L_0^\dagger = L_0, \quad L_\pm^\dagger = L_\mp, \quad P_\pm^\dagger = P_\mp^\dagger. \qquad (91)$$

Because of the tensor product structure of the representations (82), the unitarity conditions on the $\widetilde{\text{SL}(2, \mathbb{R})}$ side and on the Weil side are independent. Starting with the later, the unitarity of the representation for all $c \in \mathbb{R}\backslash\{0\}$ is assured by Stone–von Neumann's theorem. For the special linear part, the Hermiticity of $L_0$ implies that $n \in \mathbb{R}$. The vectors $|n\rangle$ can always be taken to be orthonormal, as they

---

[7]More precisely, the Weil representation is a representation of the metaplectic group $\text{Mp}(2, \mathbb{R})$, which is the double cover of $\text{SL}(2, \mathbb{R})$. The universal cover of the metaplectic group is also $\widetilde{\text{SL}(2, \mathbb{R})}$ and the representation satisfying condition (7) is $\mathring{D} = \mathring{D} \circ \pi$, where $\pi: \widetilde{\text{SL}(2, \mathbb{R})} \longrightarrow \text{Mp}(2, \mathbb{R})$ is the projection map. Since we will only write the explicit representation at the algebra level, this detail does not matter here.

are eigenvectors of a Hermitian operator. Thus, the remaining conditions can be written

$$\langle n|L_-L_+|n\rangle = (n-\Delta)(n-1+\Delta) \geq 0,$$
$$\langle n|L_+L_-|n\rangle = (n+\Delta)(n+1-\Delta) \geq 0, \quad (92)$$

for all $n$ in a given representation. These conditions are satisfied by the following classes of irreducible representations:

(i) *The trivial representation*

$$\Delta = \epsilon = 0, \quad n = 0.$$

(ii) *The continuous series representations*

$$\Delta > |\epsilon|(1-|\epsilon|), \text{ for } |\epsilon| \leq \frac{1}{2}, \quad n = \epsilon + \mathbb{Z}.$$

(iii) *The positive and negative discrete series representations*

$$\Delta > 0, \quad \epsilon = \Delta, \quad n = \Delta + \mathbb{N},$$
$$\Delta > 0, \quad \epsilon = -\Delta, \quad n = -(\Delta + \mathbb{N}).$$

We note that only the positive discrete series have a lowest weight state which makes it the obvious candidate for physical considerations.

We end this section by noting that, while the complex structure (91) is forced by the cross commutation relation and the Hermiticity of the compact generator $L_0$, one can also consider a modified version of the commutation relations where the right-hand side of every commutator is multiplied by $i$. Then, every operator can be taken to be Hermitian. The new basis of the thus obtained algebra is related to the one used above by the simple change of basis

$$\tilde{L}_0 = \frac{i}{2}(L_+ - L_-), \quad \tilde{L}_+ = L_0 - \frac{1}{2}(L_+ + L_-),$$
$$\tilde{L}_- = \left(L_0 + \frac{1}{2}(L_+ + L_-)\right),$$
$$\tilde{P}_+ = \frac{i}{\sqrt{2}}(P_+ - P_-), \quad \tilde{P}_- = \frac{1}{\sqrt{2}}(P_+ + P_-). \quad (93)$$

Since the two complex structure choices are related by a change of basis at the algebra level, the unitary irreducible representations are equivalent. In fact, this change of basis will be used in the next section to describe the gluing procedure.

### C. Gluing procedure

We are now in a position to generalize the gluing procedure of [25]. Let us start by briefly describing the motivation behind the procedure. In a quantum theory, it is well known that, given two Cauchy slices $\Sigma, \bar{\Sigma}$ with their associated Hibert spaces $\mathcal{H}_\Sigma, \mathcal{H}_{\bar\Sigma}$, the Hilbert space corresponding to the union of the spatial regions is simply the tensor product of the individual Hilbert spaces,

$$\mathcal{H}_{\Sigma \cup \bar\Sigma} = \mathcal{H}_\Sigma \otimes \mathcal{H}_{\bar\Sigma}. \quad (94)$$

In gauge theories, however, the gauge constraints relate the data from both Cauchy slices, which results in the failure of the tensor product factorization (94). Instead, the glued Hilbert space $\mathcal{H}_{\Sigma \cup \bar\Sigma}$ is a proper subset of the naive tensor product. This issue is at the core of the difficulty in defining local subregions in gravity. Splitting a diffeomorphism invariant theory into subregions introduces physical charges on the shared boundary of the distinct regions, also called the entangling surface [3]. The proposal made in [25] is simply that if $\Sigma$ and $\bar\Sigma$ share a common entangling surface, the value of the charge appearing from $\Sigma$ having a boundary should be equal to the value of the charge appearing from $\bar\Sigma$ having a boundary. In other words, the boundaries of the two regions are identified—or glued together—by matching their boundary charges. This procedure is exactly what entangles the two subregions together and the resulting entanglement entropy is the subject of upcoming work. In [25], we proposed that the charge equality condition should be applied, at the quantum level, to the maximal set of commuting operators within the charge algebra.

In the gravitational case at hand, the Hilbert space associated to a segment is completely determined by its corner. It is simply the representation spaces of the previous sections. In order to perform the gluing procedure, we start by choosing a different basis of the qcs:

$$X = \frac{C^{-1}}{\sqrt{2}}(P_+ + P_-), \quad P = \frac{i}{\sqrt{2}}(P_+ - P_-),$$
$$H = L_0 + \frac{1}{2}(L_+ + L_-), \quad K = L_0 - \frac{1}{2}(L_+ + L_-),$$
$$D = \frac{i}{2}(L_+ - L_-). \quad (95)$$

We further choose $\{\mathcal{C}^{(3)}_{\text{QCS}}, C, X, H\}$ as our maximal set of commuting operators. The constraints coming from the Casimirs force the glued states to be in the same representation. In the following, we therefore work in a single representation defined by a fixed value of $\Delta$ and $c$. To apply the condition coming from the remaining generators $X$, $H$, we need to find a basis of $\mathcal{H}^{(\Delta,c)}$ that simultaneously diagonalizes the two operators. On the Weil representation side, this was done in [25] by choosing the position eigenbasis of the harmonic oscillator,

$$|x\rangle = \sum_{k=0}^\infty \frac{1}{\sqrt{2^k k!}} \left(\frac{c}{\pi}\right)^{\frac{1}{4}} e^{-\frac{cx^2}{2}} H_k(\sqrt{c}x), \quad (96)$$

where $H_k$ is the $k$th order Hermite polynomial. We then have

$$X^{(c)}|x\rangle = x|x\rangle, \qquad (97)$$

$$H^{(c)}|x\rangle = cx^2|x\rangle. \qquad (98)$$

For the special linear part of the representation, the $X$ generator acts trivially and is thus already diagonalized in any basis. In order to diagonalize the $H$ operator, we will work with a function basis arising from a CFT-type construction. Let $\mathcal{H}^{(\Delta)}$ be the set of infinitely differentiable complex-valued functions satisfying the following asymptotic condition:

$$\psi(y) \stackrel{|y|\to\infty}{\to} \frac{1}{|y|^{2\Delta}}\left(b_\psi + \mathcal{O}\left(\frac{1}{|y|}\right)\right), \qquad (99)$$

with $b_\psi$ some constant. Then, with an appropriate choice of scalar product, the Hilbert space hosts a unitary representation of $\widetilde{SL(2,\mathbb{R})}$, on which the generators act as

$$H^{(\Delta)}\psi(y) = \partial_y\psi(y), \quad K^{(\Delta)}\psi(y) = -(y^2\partial_y + 2\Delta y)\psi(y),$$
$$D^{(\Delta)}\psi(y) = (y\partial_y + \Delta)\psi(y). \qquad (100)$$

The fact that those representations are equivalent to the irreducible representations described in the previous section is shown in [43].[8] The eigenfunctions of the $H$ operator are now simply given by the Fourier transform $\psi(q)$, whose existence is ensured by the asymptotic condition (99). We can thus write the representation space of the $\widetilde{QCS}$ as the set of vectors,

$$\psi(q) \otimes |x\rangle \equiv \Psi(q,x) \in \mathcal{H}^{(\Delta,c)}. \qquad (101)$$

The gluing procedure is then simply done as follows. Take a right and left copy of the qcs algebra and their associated Hilbert spaces. Before the constraints, the Hilbert space of the glued segment can be written

$$\tilde{\mathcal{H}}_G = \{\tilde{\Psi}_G = \Psi_L(q,x) \otimes \Psi_R(p,y)\}. \qquad (102)$$

---

[8]For the continuous series, it can be seen easily from the explicit basis

$$\psi_n(y) = \frac{1}{\sqrt{2\pi}}\left(\frac{1-iy}{1+iy}\right)^n \frac{2^\Delta}{(1+y^2)^\Delta},$$

on which we have

$$L_0^{(\Delta)}\psi_n(y) = n\psi_k(y), \quad L_-^{(\Delta)}\psi_n(y) = (n-\Delta)\psi_{n-1}(y),$$
$$L_+^{(\Delta)}\psi_n(y) = (n+\Delta)\psi_{n+1}(y).$$

The constraints

$$H_R^{(\Delta,c)}\Psi_G = H_L^{(\Delta,c)}\Psi_G, \quad X_R^{(\Delta,c)}\Psi_G = X_L^{(\Delta,c)}\Psi_G \qquad (103)$$

then simply force the arguments of the left and right functions to be the same:

$$\mathcal{H}_G = \{\Psi_G(q,x) = \Psi_L(q,x) \otimes \Psi_R(q,x)\}. \qquad (104)$$

Similarly to the special case of [25], the inverse procedure is performed by taking the state associated with a fictitious corner separating two subregions, doubling the state by tensor product with itself and finally relaxing the gluing condition. The Hilbert space (104) is the starting point of the corner proposal's approach to entanglement entropy between spatial regions. The full extent of this analysis will be explored in upcoming works.

## V. DISCUSSION AND OUTLOOK

In this paper, we used Mackey's theory of induced representations in order to fully classify the physical representations of the extended corner symmetry group. These include the case of vanishing central element in Sec. III, as well as the projective ones in Sec. IV.

Using Wigner's little group method, the former were found to be fully determined by the value of the ECS cubic Casimir. This complements the classical orbit analysis of [12]. In the Galilean case, the representations with vanishing central extension correspond to massless particles. The corresponding ECS representations could therefore be of great physical significance. It is important to note that those representations are fundamentally different from the ones with nonvanishing central element, and cannot be obtained from the latter by a limiting procedure. While the precise physical interpretation of these representations remains unclear, we leave these questions for future work and now comment on the case of nonvanishing central extension.

These representations are labeled by two parameters corresponding to the two Casimirs of the extended algebra. Although the physical significance of these quantum numbers is not completely understood, the structure of the representations may provide valuable insight. Specifically, we demonstrated that they are product representations between the $\widetilde{SL(2,\mathbb{R})}$ and Weil representations. Notably, this implies that the value the QCS Casimir on these representations is proportional to that of the special linear Casimir. In the appropriate convention, the proportionality coefficient corresponds precisely to the value of the central element. While there is no widely accepted geometric understanding of the ECS Casimir, the special linear one is believed to correspond to the corner's area operator [3,21]. It would therefore appear that reintroducing the normal translations into the discussion does not compromise the physical understanding behind the $SL(2,\mathbb{R})$ Casimir, as long as the central extension is properly taken into account. Furthermore, the central element naturally brings a scale to the system which is promising for considerations of entanglement entropy.

Another way to understand the Hilbert spaces discussed in this paper is through edge modes. At the classical level, two-dimensional gravity does not have any propagating bulk degrees of freedom. The only dynamical objects are, therefore, the edge modes living on the boundary. The gravitational edge mode is an embedding field [3,11] and thus simply corresponds to the coordinates of the corner in two dimensions. At the classical level, this explains why the coadjoint orbits are at most four dimensional. At the quantum level, the representations described in this paper can be understood as the Hilbert spaces of those edge modes.

Let us make two final remarks on these representations. First, the careful treatment of the projective representations implied that the universal cover of the special linear group should be used. This directly affects the possible values of the Casimir, and consequently the associated physical theory. For instance, in the discrete series, the parameter $\Delta$ can take any positive real value, whereas in the $SL(2, \mathbb{R})$ representations, it is restricted to half-integer and integer values. Second, only the $c > 0$ representations were considered here. The negative cases should also be included in the enumeration of the representations. As we mentioned, they are indeed not equivalent to the positive ones. For example, a negative value of the central element corresponds to an inversion of the roles of $P_-$ and $P_+$ in the Heisenberg algebra. The cross commutators with the special linear generators staying the same has direct consequences on the gluing procedure described in Sec. IV C. For more details on the representation theory of the QCS, we refer the reader to [39].

The return of the special linear representations to the forefront has another promising consequence. In the gluing procedure generalization, it was shown that the representation spaces of the QCS can be understood as one-dimensional conformal fields with an additional index in the Fock space of the harmonic oscillator. Since the symmetry group arose from considerations of two-dimensional gravity, this fact is very reminiscent of the AdS/CFT correspondence [44]. Furthermore, it has been shown that one-dimensional conformal quantum mechanics is intimately related to the causal diamond and vacuum thermal effects [45–48]. The representation theory developed here could therefore serve as the missing link between the corner proposal and the causal diamond structure. This would represent a great success for the proposal, as it would allow to get an emergent notion of spacetime geometry from the symmetries only. Furthermore, entanglement entropy is well understood within this context and, with this interpretation, could correspond to the sought-after entanglement entropy between spatial regions. These lines of inquiry will be the subject of upcoming works.

The generalization of the gluing procedure at the end of Sec. IV complements the work started in [25]. It is, in a sense, the simplest procedure one can think of working with corners. Its trivial generalization to a higher number of corners is reminiscent of the setup of causal dynamical triangulation [49]. This potential link should be investigated carefully, as it could connect the corner proposal to a top-down model of two-dimensional quantum gravity.

Another key area for exploration is the reintroduction of corner diffeomorphisms and examining how this framework extends to higher dimensions. First, one needs to understand the fate of the central extension and Casimir when tangential directions are reintroduced. One approach following [12] is to study the coadjoint orbits of the QCS, before reintroducing the diffeomorphisms using a Lie algebroid structure. Another interesting avenue lies in the semidirect product structure between the diffeomorphisms and the finite dimensional corner symmetry group. This suggests a "little group approach" to representation theory, where the little group would consist of the subset of diffeomorphisms that leave a particular representation of the corner symmetries invariant. In the case of the ECS, that would be the subset of corner diffeomorphisms leaving the cubic Casimir (36) invariant. In the case of the QCS, it would correspond to the subset of corner diffeomorphisms for which the cubic Casimir (75) and the central extension are invariant. By finding the irreducible representations of these little groups, one can then induce them to the full group using Mackey's method, potentially yielding irreducible representations of the full corner symmetry group in higher dimensions.

## ACKNOWLEDGMENTS

I am grateful to Jerzy Kowalski-Glikman for helping me navigate the different topics of this paper and helping me decide what was important and how to direct my research. Many thanks to Luca Ciambelli for putting up with my ramblings about representation theory and guiding me through it. I have greatly benefited from discussions with Jackie Caminiti, Federico Capeccia, José Figueroa-O'Farrill, Laurent Freidel, and Rob Leigh. I thank Perimeter Institute for their hospitality.

## DATA AVAILABILITY

No data were created or analyzed in this study.